**Global Rice Multi-Class Segmentation Dataset (RiceSEG): A Comprehensive and Diverse High-Resolution RGB-Annotated Images for the Development and Benchmarking of Rice Segmentation Algorithms**

Junchi Zhou[1], Haozhou Wang[2], Yoichiro Kato[2], Tejasri Nampally[3], P. Rajalakshmi[4], M. Balram[5], Keisuke Katsura[6], Hao Lu[7], Yue Mu[1], Wanneng Yang[8], Yangmingrui Gao[1], Feng Xiao[1], Hongtao Chen[1], Yuhao Chen[1], Wenjuan Li[9], Jingwen Wang[10], Fenghua Yu[11], Jian Zhou[12], Wensheng Wang[13], Xiaochun Hu[14], Yuanzhu Yang[14], Yanfeng Ding[1], Wei Guo[2, *], Shouyang Liu[1, *]

1 Engineering Research Center of Plant Phenotyping, Ministry of Education, Jiangsu Collaborative Innovation Center for Modern Crop Production, Academy for Advanced Interdisciplinary Studies, Sanya Institute of Nanjing Agricultural University, Nanjing Agricultural University, Nanjing, China.

2 Graduate School of Agricultural and Life Sciences, The University of Tokyo, Tokyo, Japan

3 Department of Artificial Intelligence, Indian Institute of Technology, Hyderabad, India

4 Department of Electrical Engineering Indian Institute of Technology, Hyderabad, India

5 Institute of Biotechnology, Professor Jayashankar Telangana Agricultural State University, Hyderabad, India

6 Graduate School of Agriculture, Kyoto University, Kyoto, Japan

7 Key Laboratory of Image Processing and Intelligent Control, School of Artificial Intelligence and Automation, Huazhong University of Science and Technology, Wuhan, China

8 National Key Laboratory of Crop Genetic Improvement, National Center of Plant Gene Research, and Hubei Key Laboratory of Agricultural Bioinformatics, Huazhong Agricultural University, Wuhan, China

9 State Key Laboratory of Efficient Utilization of Arid and Semi-arid Arable Land in Northern China, the Institute of Agricultural Resources and Regional Planning, Chinese Academy of Agricultural Sciences, Beijing, China.

10 Center for Geospatial Information, Shenzhen Institutes of Advanced Technology, Chinese Academy of Science, Shenzhen, China.

11 School of Information and Electrical Engineering, Shenyang Agricultural University, Shenyang, China

12 Rice Research Institute, Jilin Academy of Agricultural Sciences, Changchun, China

13 Institute of Crop Sciences/National Key Facility for Crop Gene Resources and Genetic Improvement, Chinese Academy of Agricultural Sciences, Beijing, China

14 Yuan Long Ping High-Tech Agriculture Co., Ltd., Changsha, China

**\*Corresponding authors:**

**Shouyang Liu, shouyang.liu@njau.edu.cn;**

**Wei Guo, guowei@g.ecc.u-tokyo.ac.jp**

Abstract:

Developing computer vision–based rice phenotyping techniques is crucial for precision field management and accelerating breeding, thereby continuously advancing rice production. Among phenotyping tasks, distinguishing image components is a key prerequisite for characterizing plant growth and development at the organ scale, enabling deeper insights into eco-physiological processes. However, due to the fine structure of rice organs and complex illumination within the canopy, this task remains highly challenging, underscoring the need for a high-quality training dataset. Such datasets are scarce, both due to a lack of large, representative collections of rice field images and the time-intensive nature of annotation. To address this gap, we established the first comprehensive multi-class rice semantic segmentation dataset, RiceSEG. We gathered nearly 50,000 high-resolution, ground-based images from five major rice-growing countries (China, Japan, India, the Philippines, and Tanzania), encompassing over 6,000 genotypes across all growth stages. From these original images, 3,078 representative samples were selected and annotated with six classes (background, green vegetation, senescent vegetation, panicle, weeds, and duckweed) to form the RiceSEG dataset. Notably, the sub-dataset from China spans all major genotypes and rice-growing environments from the northeast to the south. Both state-of-the-art convolutional neural networks and transformer-based semantic segmentation models were used as baselines. While these models perform reasonably well in segmenting background and green vegetation, they face difficulties during the reproductive stage, when canopy structures are more complex and multiple classes are involved. These findings highlight the importance of our dataset for developing specialized segmentation models for rice and other crops. The RiceSEG dataset is publicly available at www.global-rice.com.

# 1.Introduction

As a core pillar of global agricultural production, rice is widely cultivated worldwide that feeds more than half of the global population (Jin et al., 2020). Yet, facing the global warming, the variability and uncertainty in rice growing environments pose severe challenges for the sustainability of rice production (Godfray et al., 2010). To leverage the unfavorable growth conditions, great efforts have been made to improve both the cultivars and cultivation practices, according to the adaptation of phenotypic traits (Cassman et al., 1995). Hence, the success of these efforts deeply relies on the precision and throughput of the plant phenotyping techniques. Unfortunately, the measurement of plant phenotypic traits is mainly accomplished manually that are very time-consuming and labor-intensive (C. Chen et al., 2006; Madec et al., 2019; Mandal et al., 2018). Developing high-throughput phenotyping techniques is therefore crucial to overcome the limitations and consequently ensure the rice production (Maohua, 2001; Mermut et al., 2001; Yandun Narvaez et al., 2017).

Compared with traditional human observation, computer vision techniques have greatly advanced plant phenotyping by providing higher throughput and accuracy (Z. Li et al., 2020). A key step in this domain is image segmentation, which underpins the extraction of critical traits such as canopy structure (J. Wang et al., 2020), light interception (Shouyang et al., 2020), and stress status (Baret et al., 2018). For single-class segmentation, distinguishing green vegetation from the background, deep learning-based models have demonstrated robust performance across various crops (Gao et al., 2023a; Serouart et al., 2022), maintaining consistent accuracy under diverse environments, genotypes, and spatial resolutions (Gao et al., 2024a). However, there is a growing need for more detailed segmentation that distinguishes multiple plant organs (e.g., panicles, and both green and senescent leaves), as this enables deeper insights into organ development and the source–sink relationship (Z. Zhao et al., 2022). Moreover, because weeds commonly appear in rice fields, simultaneously segmenting weeds alongside crop organs both reduces misclassification and informs weed management strategies. Although recent deep learning segmentation models, such as SAM (Kirillov et al., 2023), show promise, none have successfully addressed the multi-class segmentation of rice canopies, encompassing both organs and weeds, across diverse genotypes and environmental conditions. This is primarily due to the unique challenges posed by rice canopies, which feature fine leaves, thin stems, and substantial genotype-dependent variations. Fluctuating field illumination further complicates segmentation by creating mutual shading within the canopy, while reflective water surfaces in paddy fields produce mirror-like reflections and glare, distorting certain image regions and reducing clarity. As with other complex computer vision tasks, improving current models or developing specialized approaches hinges on the availability of comprehensive training datasets that capture the full complexity of rice field conditions.

High-quality training datasets are critical for adapting state-of-the-art computer vision (CV) models to plant phenotyping (Garcia-Garcia et al., 2017). In recent years, numerous phenotyping datasets have emerged for various crops, both indoors and in the field, focusing primarily on plant counting (Bai et al., 2023), organ detection (David et al., 2020, 2021), and disease or pest classification (Prajapati et al., 2017; Wu et al., 2019). However, few datasets target semantic segmentation due to the labor-intensive nature of pixel-level annotation. This issue is especially pronounced in rice, where fine leaves and dense canopies complicate the annotation process, leading to a shortage of publicly available datasets (Cordts et al., 2016; Russell et al., 2008). Table 1 provides an overview of representative plant semantic segmentation datasets, which for rice crops are largely confined to single classes—either panicle segmentation (H. Wang et al., 2021)for basic green segmentation from the background (Madec et al., 2023). In summary, no existing rice segmentation dataset jointly encompasses multiple genotypes, diverse field conditions, multiple organs (leaf, stem, and panicle), and weeds.

**Table 1** Representative semantic segmentation datasets.

| Dataset | Crop type | Class | # Images | Image Size |
|---|---|---|---|---|
| CVPPP(Scharr et al., 2014) | Rosette plants | 2 | 1311 | 2048 × 2448 |
| CWFID(Haug et al., 2015) | Carrot | 3 | 60 | 1291 × 966 |
| Oil Radish Growth (Mortensen et al., 2019) | Oil radish | 7 | 129 | 1601 × 1601 |
| PhenoBench(Weyler et al., 2023) | Sugar beet | 3 | 2872 | 1024 × 1024 |
| Paddy Rice Imagery (H. Wang et al., 2021) | Rice | 2 | 400 | 4096 × 2160 |
| VegAnn (Madec et al., 2023) | Rice, wheat etc. | 2 | 466 | 512 × 512 |
| RiceSEG | Rice | 6 | 3078 | 512 × 512 |

The main objectives of this work would build a broad, multi-class, high-resolution semantic segmentation dataset for rice crops. This dataset includes 3078 ground-based RGB images collected from 5 countries and 12 different institutions, taken along the whole growth cycle, and covering wide range of genotype-environment-management combinations. Pixels in all the images are finely annotated into six categories: background, green vegetation, senescent vegetation, panicle, weed, and duckweed. Furthermore, to assess the dataset, we also report baseline results for the most classic and cutting-edge semantic segmentation algorithms. The main contributions of this study are twofold:

- To the best of our knowledge, we present the largest global rice semantic segmentation dataset, offering precise pixel-level annotations across multiple detailed classes in real rice fields.
- We conducted extensive experiments with various segmentation models on this dataset to establish benchmark performance, thereby facilitating the development of more effective rice segmentation algorithms.

## 2. Materials and Methods

### 2.1 Dataset collection

To maximize the representativeness and diversity of the dataset, we collected approximately 50,000 images in total, contributed by 12 institutions between 2012 and 2024, from 14 sites located in 5 countries, including China, Japan, India, the Philippines, and Tanzania (Table 2). They were taken by different types of cameras, such as Digital Single-Lens Reflex cameras, portable action cameras, or smartphones. The configuration of the cameras was set 1-2 m above the canopy with different orientations (0º - 90º) towards the canopy. This ensures the high-resolution of the images with the ground sampling distance (GSD) ranging from 0.1-1.8 mm/pixel.

**1) Dataset from China.** The dataset originates from various sites across China, encompassing all major rice production regions from the northeastern most to the southern most areas where rice is cultivated. This extensive coverage includes over 6000 rice varieties, resulted in a large collection of diverse images. Specifically, images provided by Nanjing Agricultural University (JS_1, JS_2, JS_3, JS_4, HN) were meticulously gathered from experimental fields in Jiangsu and Hainan provinces, featuring over one thousand rice varieties. These images highlight the challenges of segregating plant organs due to the high variability in canopy structures among genotypes under diverse field light conditions, as well as the presence of weeds or duckweed in the background. Additionally, images from Changsha were captured in the rice experimental fields of Yuan Long Ping High-Tech Agriculture Co., td (https://lpht.com.cn/), a leading firm in rice breeding renowned for its hybrid rice varieties. This collection includes images of nearly 5,000 rice genotypes at various growth stages (transition and reproductive stage), encompassing both domestic and international varieties.

The northeastern region significantly contributes to China's rice production, particularly known for high-quality japonica rice adapted to cold climates. Images were collected from each of the northeastern provinces, including Heilongjiang (HL), Jilin (JL), and Liaoning (LN). The 'HL' dataset was captured by Institute of Agricultural Resources and Regional Planning using a fisheye camera, providing a unique wide-angle perspective of the rice canopy across several varieties. The 'JL' dataset comprises images from over 700 rice varieties obtained from the Rice Research Institute of the Jilin Academy of Agricultural Sciences, while the 'LN' dataset was provided by Shenyang Agricultural University. The 'JX' and 'GX' sub-datasets, contributed by Huazhong University of Science and Technology, document images from various growth stages ranging from seedling to jointing across more than 40 genotypes in Jiangxi and Guangxi provinces, respectively. The 'HB' sub-dataset, provided by Huazhong Agricultural University, includes data from 104 varieties, and the 'GD' dataset, supplied by the Shenzhen Institute of Advanced Technology, Chinese Academy of Sciences, encompasses images from over 60 genotypes.

**2) Dataset from Japan.** This dataset encompasses a broad spectrum of rice genotypes in Japan. Notably, the dataset sourced from the University of Tokyo (TKO_1, TKO_2, TKO_3) comprises time-series images of rice captured by field-fixed cameras. The UTokyo dataset was collected from paddy phenotyping field trials at the Institute for Sustainable Agro-ecosystem Services (ISAS) 35°44'20.3"N 139°32'29.8"E) in Tokyo, Japan, in the 2014 season. A Field Server system ¥cite{utokyo_2015} collected images of five genotypes through the whole growth stage. The camera module of the system is based on a digital single-lens reflex (DSLR) camera, the Canon EOS Kiss X5 camera, with an EF-S18-55 mm lens (Canon Inc., Tokyo) that provides high-quality and high-resolution (18 megapixels) image data. A preprogrammed microcontroller board controls the power and shutter of the camera automatically.

**3) Dataset from India.** The dataset is obtained from Institute of Biotechnology of Professor Jayashankar Telangana State Agriculture University, located in Hyderabad, Telangana, India. The study area covers an area of 15.3 m x 34.8 m and includes two repetitions of 203 plots, each representing a different variety/genotype of aerobic paddy, resulting in a total of 406 plots. Each plot covers an area of 1.26 square meter and contains 42 crop strands. The dataset provides a collection of images of upland rice, which are unique due to the presence of many weeds in complex backgrounds. The images were captured by a team from the Indian Institute of Technology Hyderabad using a high-resolution Sony RX 100 camera. Each image is of resolution of 3456 x 2592 pixels.

**4) Dataset from Philippines.** The dataset was collected from the International Rice Research Institute (IRRI) farm located in Los Baños, Philippines, at coordinates 14°11 N, 121°15 E, and an elevation of 21 meters above sea level. The study encompasses three distinct paddy fields containing a comprehensive collection of rice varieties with varying experimental conditions. In total, the dataset comprises 1596 rice varieties/lines distributed across 2172 plots, with some overlap in varieties between fields. All fields maintained a consistent planting density of 20 cm x 20 cm spacing between plants, creating a uniform growing environment for comparative analysis. The experimental design allows for systematic evaluation of rice phenotypes under different field management strategies. All images were captured during the vegetative stage of rice growth, specifically 3-4 weeks after transplanting.

**5) Dataset from Tanzania.** Field experiments were conducted at the irrigated lowland field in Kilimanjaro Agricultural Training Centre in Republic of Tanzania (3°45'08" S, 37°39'68'' E, 720 m above sea level) in 2019. Four rice varieties, NERICA 1, IR64, TXD 306 and Wahiwahi, were four different water managements with three replications: continuous flooded condition, three alternate wetting and drying conditions, irrigation was repeated until the water depth reached 10 cm when the surface water level dropped to 0 cm 15 cm and 30 cm, respectively. At maturity, images of the rice canopy were taken vertically downward using a digital camera (WG-4, Ricoh, Japan) from 80 cm above the rice canopy. Twenty-four rice hills (4 hills × 6 hills, 1.2 m × 0.9 m) that were captured in the images were then harvested at ground level and the yield and yield components were investigated.

**Table 2** Metadata of the sub-datasets comprising the RiceSEG Dataset

| | Name | Institute | Site | Images | Lat (°) | Long (°) | Year | Growth stage[a] | Genotypes | Platform | Camera | Image size (pixels) | GSD (mm/px) |
|---|---|---|---|---|---|---|---|---|---|---|---|---|---|
| CHINA | JS_1 | NJAU | Jiangsu | 4000 | 31.5 N | 119.3 E | 2020 | Vegetative, Transition | 1000 | Handheld rod | SONY RX0 | 4800*3200 | 0.1-0.3 |
| | JS_2 | | | 4000 | | | 2021 | Vegetative | | | | | |
| | JS_3 | | | 8000 | | | 2023 | Vegetative, Transition | | | | | |
| | JS_4 | | | 8000 | | | 2023 | Reproductive | | | | | |
| | HN | | Hainan | 2000 | 18.2 N | 109.5 E | 2023 | Vegetative, Transition | | | | | 0.3-0.5 |
| | GX | HUST | Guangxi | 280 | 24.3 N | 109.4 E | 2012 | Vegetative | 20 | Fixed rod | Canon EOS 1100D | 4272*2848 | 0.3&1.2 |
| | JX | | Jiangxi | 355 | 28.7 N | 115.9 E | 2013 | Vegetative | 35 | Fixed rod | OLYMPUS E-450 | 3648*2736 | 1.8 |
| | HB | HZAU | Hubei | 104 | 30.5 N | 114.3 E | 2016 | Transition | 104 | Tripod | NIKON D7100 | 6000*4000 | 0.3 |
| | HL | CAAS | Harbin | 40 | 45.7 N | 126.6 E | 2016 | Vegetative | 40 | Handheld rod | NIKON D7100 | 2000*2000 | 0.6 |
| | GD | CAS | Guangdong | 90 | 22.6 N | 113.1 E | 2022 | Reproductive | 60 | Handheld rod | iphone11 | 2048*1536 | 0.1-0.3 |
| | LN | SYAU | Shenyang | 154 | 41.8 N | 123.4 E | 2024 | Vegetative, Transition | 50 | Handheld rod | SONY RX0 | 4800*3200 | 0.1-0.3 |
| | HUN | LPHT | Changsha | 14994 | 28.2 N | 112.9 E | 2024 | Transition | 5000 | Handheld rod | SONY RX0 | 4800*3200 | 0.1-0.3 |
| | JL | JAAS | Changchun | 2642 | 43.8 N | 125.3 E | 2024 | Reproductive | 700 | Handheld rod | SONY RX0 | 4800*3200 | 0.1-0.3 |
| JAPAN | TKO_1 | UTokyo | Tokyo | 645 | 35.4 N | 139.3 E | 2013 | Vegetative | 5 | Fixed rod | Canon EOS Kiss x5 | 5184*3456 | 0.1 |
| | TKO_2 | | | 142 | | | 2014 | All stage | | | | | |
| | TKO_3 | | | 768 | | | 2015 | Transition | | | | | |
| INDIA | TG | IITH | Telangana | 271 | 17.3 N | 78.4 E | 2018 | Vegetative | 50 | Handheld rod | Sony RX100 | 5472*3648 | 0.3-0.5 |
| TANZANIA | Kil | KU | Kilimanjaro | 126 | 3.45 S | 37.4 E | 2019 | Reproductive | 4 | Handheld rod | RICOH WG-4 | 3072*2304 | 0.2-0.4 |
| PHILIPPINES | Lag | IRRI | Laguna | 200 | 14.2 N | 121.2 E | 2014 | Vegetative | 1596 | Handheld rod | OLYMPUS TG-620 | 1600*1200 | 0.3-0.5 |

[a]Growth stages of rice, categorized into three main phases: (a) Vegetative: Seedling, Tillering, and Jointing; (b) Transition: Booting, Heading, Flowering; (c) Reproductive: Filling and Maturity.

## 2.2 Establishment of the RiceSEG dataset

Considering the substantial variation in the number of images collected from China and other countries (Tables 2 and Fig.1), we employed distinct selection strategies to maximize the dataset's representativeness (Fig 2 and Table 3). In China, collaborations across all major rice-growing regions enabled the largest overall collection of images compared to other countries. From each Chinese site, 60–100 images were randomly chosen to capture diverse growth stages, varieties, and environmental conditions. In contrast, acquiring high-resolution rice images from other countries proved more challenging; hence, for the remaining five countries, we utilized nearly all the originally collected data.

After finalizing image selection across all sites, a cropping procedure was adopted. With continual advancements in computational resources, larger models can leverage higher-resolution images for potentially enhanced performance (Jia et al., 2023). Nonetheless, balancing annotation costs with the demand for high-resolution imagery led us to fix the final cropping size at 512 × 512 pixels. For the Chinese dataset, a single 512 × 512 sub-image was extracted from the center of each selected image, while for images from other countries, a 1024 × 1024 region was first cropped from the center and then subdivided into one to four sub-images using a sliding-window approach, with each sub-image carefully inspected for quality.

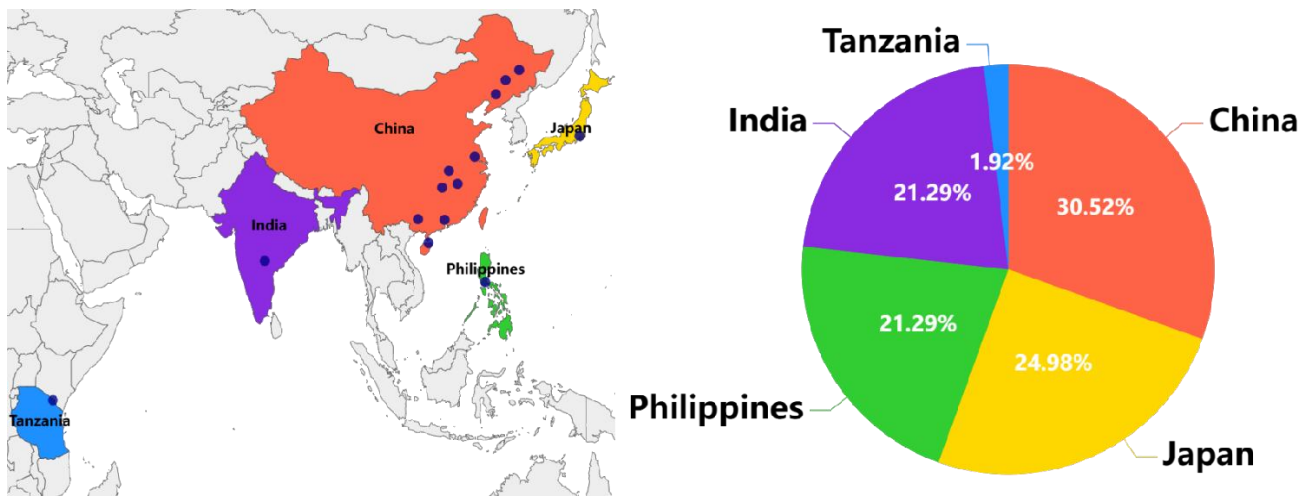

Figure 1. Global range distribution and composition of the datasets

## 2.3 Data Annotation

We engaged specially trained volunteers, primarily graduate students studying agronomy in Nanjing Agricultural University, to manually annotate the images. In total, the annotation process involved 11 volunteers, with the time cost for each image ranging from 0.5 to 1.5 hours, depending on its complexity. Collectively, the annotators dedicated 2440 hours to data annotation and an additional 800 hours to verification and refinement, culminating in a total of 3240 hours.

The training program encompassed fundamental knowledge of rice growth physiology, equipping annotators to identify diverse characteristics and morphological traits of rice at various growth stages. Participants were further trained to categorize each pixel into one of six predefined classes: background, green vegetation, senescent vegetation, panicle, weed, and duckweed, labeling as number from 0 to 5, correspondingly. The detailed explanations and annotation samples of each category was provided to ensure a consistent classification criterion (Figure 2). Moreover, annotators were trained to use a JavaScript-based image annotation tool (https://github.com/kyamagu/js-segment-annotator) (Tangseng et al., 2017). This tool was selected because it was developed based on the superpixel annotation method. This significantly enhances annotation efficiency while ensures precise alignment with natural boundaries. Note that annotators were required to adjust the superpixel resolution carefully to capture fine details and textures in rice images.

To ensure annotation quality and consistency, a strict protocol was followed throughout the process. After the initial round of annotation, approximately 10% of the labeled images from each annotator were randomly selected for double-checking by a second annotator. During this process, common misclassifications were identified and corrected, with documentation provided by the project leader. Feedback was then promptly given to the annotators to improve their practices. In summarizing the lessons learned from this iterative annotation process, we found that among the six categories, distinguishing senescent leaves, particularly those at the bottom of the canopy with substantial shadow, was often challenging. Additionally, residual plant matter from previous crop rotations sometimes resembled senescent rice, further complicating the labeling task. To minimize subjectivity, each annotation was cross-verified by at least three individuals to ensure reliability. Finally, weeds such as water onions, which structurally resemble rice at certain growth stages, were sometimes misclassified as green vegetation. Extra care was taken to maintain precision in the annotations.

Due to the nature of agricultural ecosystems, the labels in the RiceSEG dataset are not evenly distributed across categories, as expected (Table 3). The background category is the most dominant, accounting for nearly 50% of all labels. Following this, the green vegetation category ('green_veg') ranks second, comprising over 40%, as green plants cover a significant portion of the rice fields and are the primary visual component throughout the growth cycle. In contrast, categories such as senescent vegetation ('senescent_veg') and rice panicle ('panicle') only appear during the reproductive stage and thus represent a relatively small proportion of the dataset. Additionally, due to the use of herbicides across all experimental sites, the presence of weeds and duckweed is minimal.

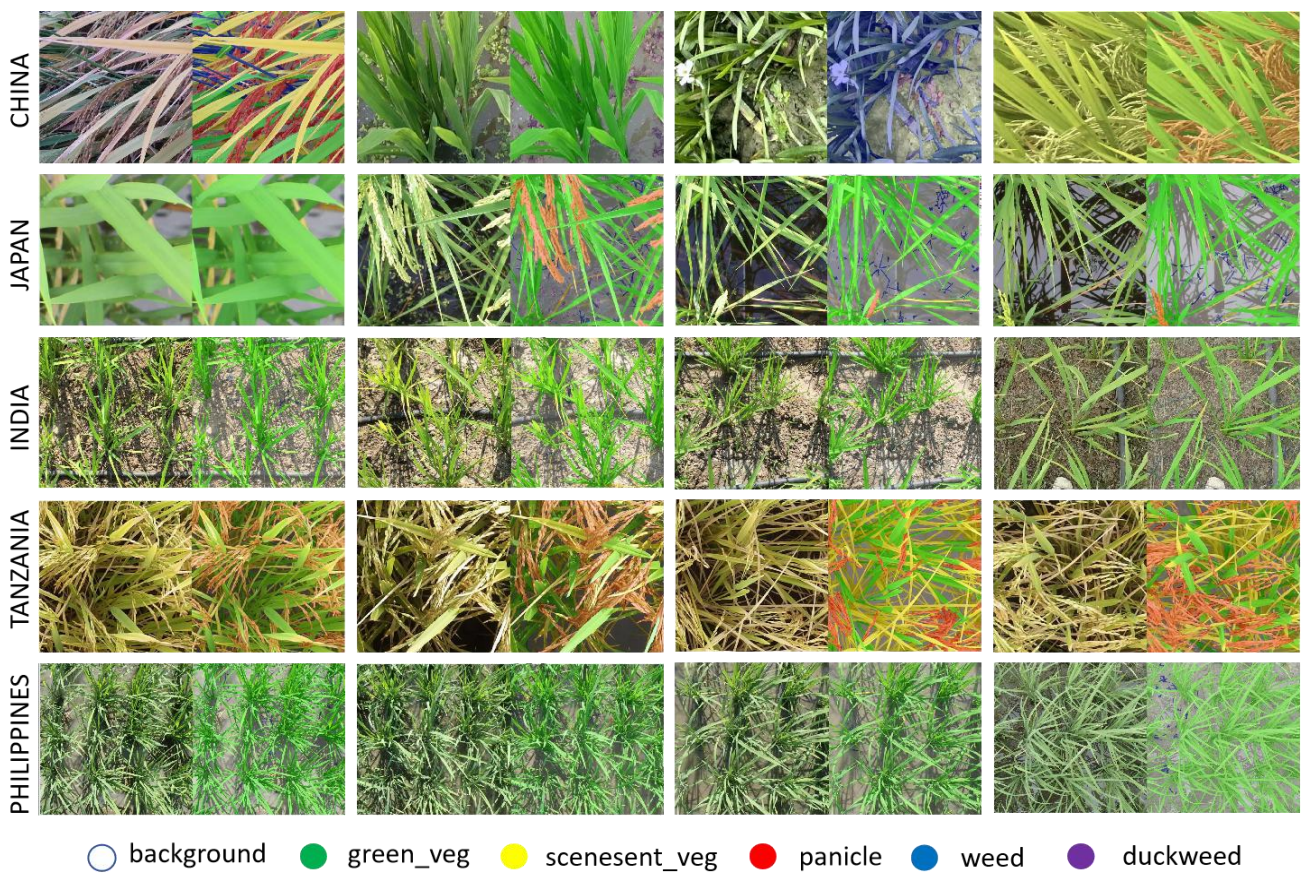

○ background ● green_veg ● scenesent_veg ● panicle ● weed ● duckweed

**Figure 2. Image Annotation Process**

Table 3 Statistics of the RiceSEG dataset

| Name | Images | No. of classes | Category Proportions (%) | | | | | |
|---|---|---|---|---|---|---|---|---|
| | | | background | green_veg | senescent_veg | panicle | weed | duckweed |
| JS_1 | 100 | 4 | 47.1 | 50.6 | 0.2 | | | 1.6 |
| JS_2 | 100 | 5 | 53.9 | 44.4 | 0.2 | 0.4 | | 1.6 |
| JS_3 | 100 | 6 | 25.3 | 22.9 | 0.4 | 0.4 | 36.2 | 14.9 |
| JS_4 | 80 | 5 | 11.4 | 37.1 | 32.7 | 16.4 | 2.8 | |
| HN | 100 | 5 | 56.8 | 42.6 | 0.2 | | 0.1 | 0.3 |
| GX | 60 | 4 | 79.0 | 19.0 | 2.1 | | 0.7 | |
| JX | 60 | 3 | 49.6 | 47.6 | | 1.9 | | |
| HB | 100 | 4 | 24.9 | 67.2 | 5.6 | 2.4 | | |
| HLJ | 100 | 2 | 48.6 | 51.5 | | | | |
| GD | 100 | 4 | 8.30 | 72.5 | 4 | 15.1 | | |
| LN | 60 | 6 | 21.5 | 70.7 | 3.6 | 1.6 | 0.2 | 2.5 |
| HUN | 100 | 5 | 3.0 | 66.4 | 4.5 | 26.1 | 0.1 | |
| JL | 60 | 5 | 6.44 | 63.2 | 1.2 | 26.4 | | 2.8 |
| TKO_1 | 100 | 4 | 49.6 | 47.6 | 2.1 | | 0.7 | |
| TKO_2 | 504 | 6 | 67.4 | 28 | 0.8 | 2.3 | 0.8 | 0.8 |
| TKO_3 | 100 | 4 | 15.4 | 82.3 | 2.1 | 0.2 | | |
| TG | 600 | 5 | 58.9 | 39.1 | 1.2 | 0.2 | 0.6 | |
| Kilimanjaro | 54 | 4 | 15.7 | 27.5 | 25.8 | 30.9 | | |
| Laguna | 600 | 4 | 55.7 | 43.5 | 0.3 | | 0.3 | |
| **Summary** | **3078** | **6** | **48.3** | **43.4** | **2.5** | **3.4** | **1.6** | **0.8** |

### 2.4 Baseline test

### 2.4.1 Baseline models

To establish the baseline accuracy for the RiceSEG dataset, we offer baseline results for six semantic segmentation models divided into two major categories (Tabel 4): Convolutional Neural Networks (CNN) and Transformer-based models. We have chosen, FCN (Long et al., 2015; Shelhamer et al., 2017), PSPNet (H. Zhao et al., 2017), and DeepLabV3+ (L.-C. Chen et al., 2018), three methods based on the CNN backbone. Regarding the Transformer architecture, we adopted SegFormer (Xie et al., 2021), KNet (Zhang et al., 2024) and Mask2Former (Cheng et al., 2022). These models represent the classic and cutting-edge technologies in the semantic segmentation field. Our RiceSEG dataset were randomly split 8:2 for the training and test dataset. All the six models selected were trained and test accordingly.

Table 4. Baseline model for semantic segmentation

| | Model | Backbone | Venue | Key Features |
|---|---|---|---|---|
| CNN backbone | FCN | Resnet50 | 2015-CVPR/2017-TPAMI (Long et al., 2015; Shelhamer et al., 2017) | Fully convolutional network for semantic segmentation |
| | PSPNet | | 2017-CVPR (H. Zhao et al., 2017) | Employing pyramid pooling to capture multi-scale contextual information. |

| | | | |
|---|---|---|---|
| | DeepLabV3+ | | 2018-ECCV (L.-C. Chen et al., 2018) | Combining atrous convolutions with a new decoder for enhanced boundary delineation. |
| Transformer backbone | SegFormer | Mit b0 | 2021-NeurIPS (Xie et al., 2021) | Efficient transformer-based model with a lightweight MLP decoder. |
| | KNet | SwinT | 2021-NeurIPS (Zhang et al., 2024) | Uses kernel-based convolution for multi-scale feature extraction. |
| | Mask2Former | | 2022-CVPR (Cheng et al., 2022) | Unifies semantic and instance segmentation with dynamic mask prediction. |

### 2.4.2 Evaluation metrics

At the pixel scale, to evaluate the baseline models, we report the Intersection over Union (IoU) and Accuracy for each class, while using the Mean Intersection over Union (mIoU) and Mean Accuracy (mAcc) as performance metrics across all classes. Then, at the image scale, we calculate the proportion of each class in the entire image and compare it with the corresponding proportions from the manually labeled images. Further, we calculate $R^2$ and RMSE to assess the model's performance.

## 3. Results

### 3.1. Dataset Diversity Analysis

Figure 3 shows the distribution pattern of the rice dataset across five countries. Overall, the data from China exhibits a relatively larger distribution area due to the broad variation in genotype-environment-management factors in the rice images collected. In contrast, the distribution of samples from the other four countries largely overlaps with the Chinese dataset, but within a narrower domain. Nevertheless, the datasets from these four countries demonstrate distinct distribution patterns. Ultimately, the combined samples from all five countries contribute to expanding the dataset's distribution and improving its representation of the diverse range of high-resolution rice field images.

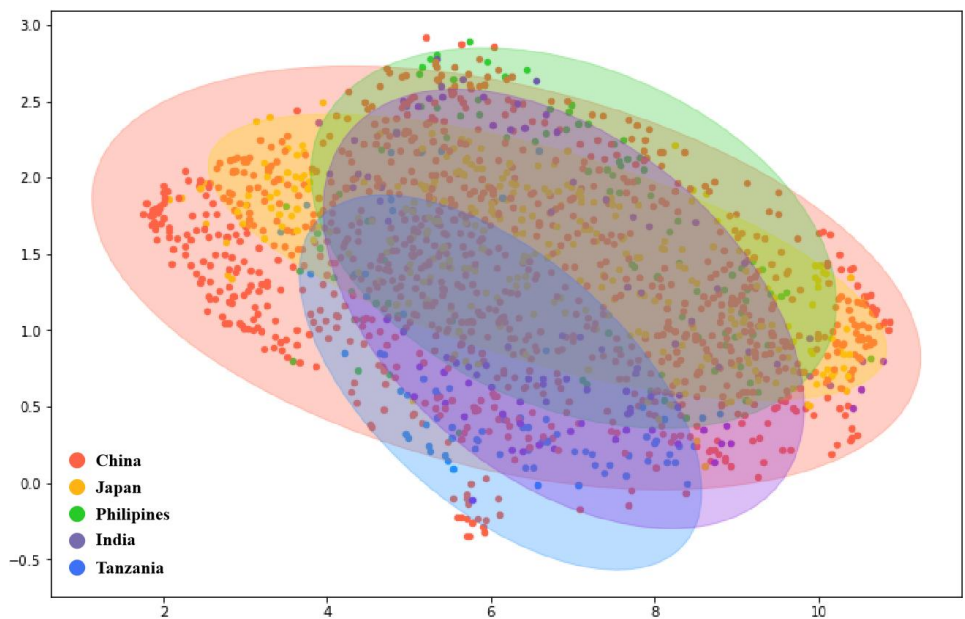

**Figure 3. Distribution of sub-datasets from different contributes. The images were projected into two dimensions using uniform manifold approximation and projection (UMAP). The points are colorized by the country where the data was collected. Each colored domain represents the confidence ellipse of the country 's dataset.**

## 3.2. Baseline results at pixel scale

Regarding the average performance across all classes, transformer-based segmentation models outperform their CNN counterparts (Table.5). Specifically, all baseline models generally perform well in segmenting background, green vegetation, and panicles. However, significant differences are observed in more challenging categories such as senescent vegetation, weeds, and duckweed. For senescent vegetation, none of the models delivered satisfactory results, with the best-performing model, Mask2Former, achieving an IoU of only 52.98, and SegFormer achieving an ACC of 66.47. For weeds, although the top-performing model reached a classification accuracy of 77.06, the IoU remained low at 65.73.

Table 5. Performance of different models on the RiceSEG.

| Metrics | CNN backbone | | | Transformer backbone | | |
|---|---|---|---|---|---|---|
| | FCN | PSPNet | DeepLabv3+ | SegFormer | KNet | Mask2Former |
| mIoU | 54.82 | 68.16 | 65.93 | 72.70 | 71.87 | **74.69** |
| mAcc | 61.85 | 80.48 | 79.35 | 83.57 | 80.50 | **83.85** |

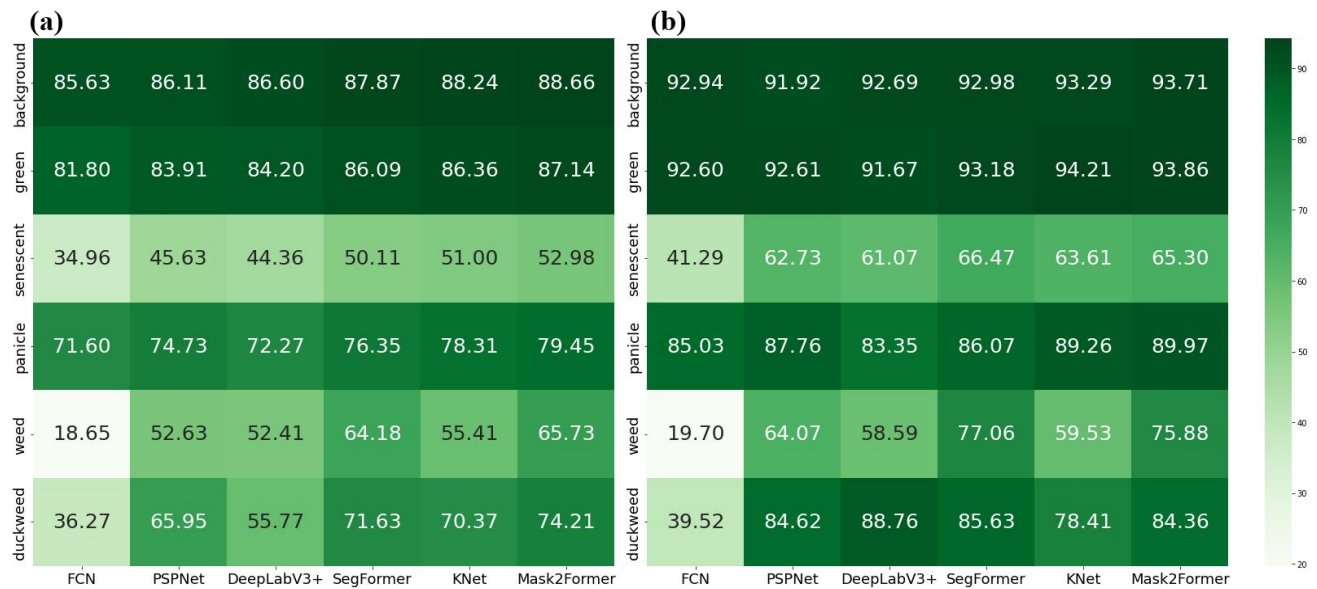

Figure 4. Segmentation performance at pixel scale. There are six classical and state of art semantic segmentation models were compared in terms of terms of IOU (a) and ACC (b). The test dataset includes 601 images for 6 classes

Figure 5 illustrates the segmentation performance of all baseline models on the test set. During the vegetative growth stages, the majority of the images consist of green vegetation and background, which are highly contrasted and easily distinguishable. As a result, only minor differences in segmentation performance were observed among the models during this phase. However, during the transition phase, segmentation becomes more challenging due to the emergence of weeds and duckweed. The high morphological similarity between these and rice parts leads to misidentification as rice, resulting in less accurate segmentation and an increased occurrence of false positives. In the reproductive stage, the canopy begins to saturate, leaving only a small portion of the background visible. This leads to the misclassification of yellow leaves, which are predominantly classified as green vegetation or background. Achieving reliable recognition performance remains difficult for both traditional CNN models and state-of-the-art transformer models.

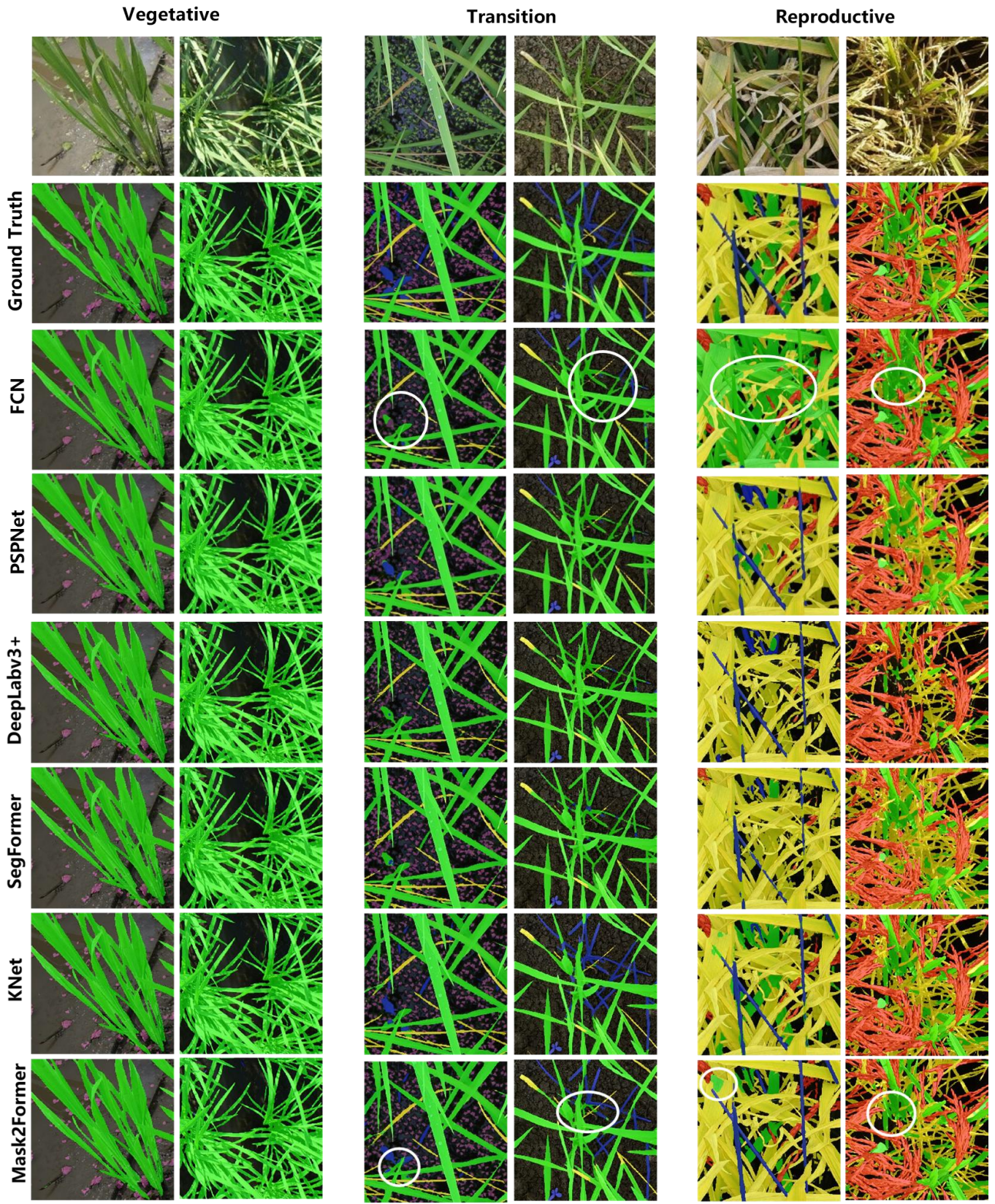

Figure 5. Visualization of the segmentation results on the test set. There are six classical and state of Art semantic segmentation models were tested on three growth stages of rice. The white circles represent key misclassified regions.

## 3.3. Baseline results at image scale

At image scale, for green vegetation and panicles, the models generally performed well. However, for more complex categories such as weeds and senescent vegetation, CNN models performed poorly. In contrast, transformer-based models significantly improved performance. Furthermore, we demonstrated the dynamics of rice canopy from seedling to maturity stage based on the best-performing Mask2Former model (Figure 7). It further indicates that the disperses of the segmentation at reproductive stage consist with that at pixel scale.

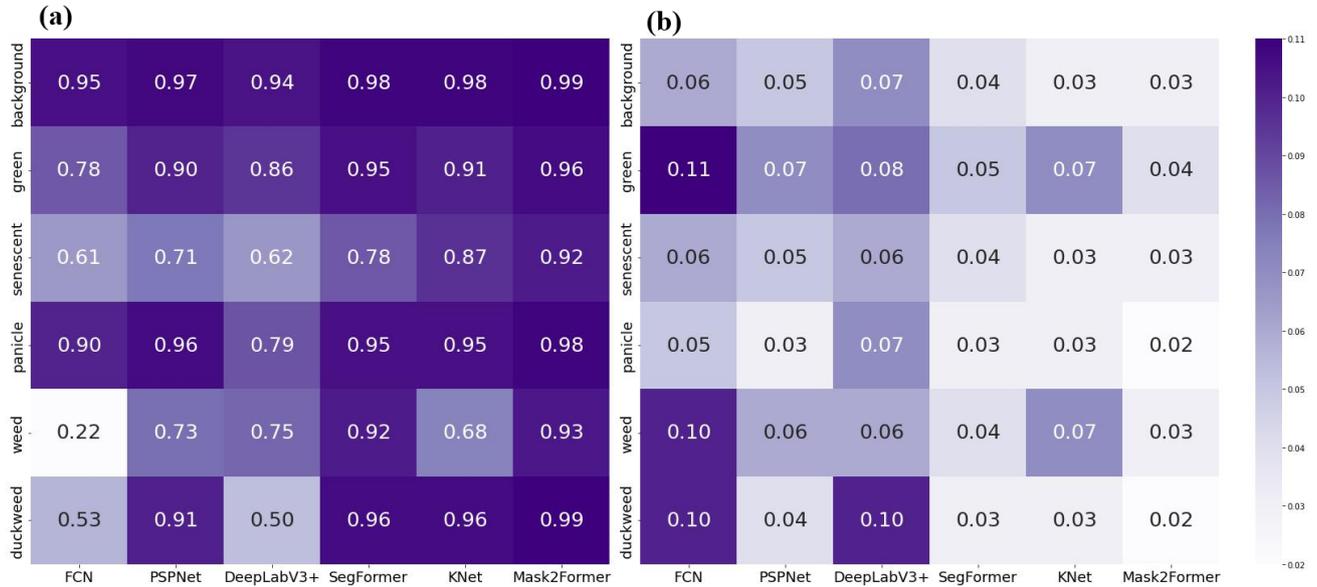

Figure 6. Segmentation performance at image scale. There are six classical and state of art semantic segmentation models were compared in terms of terms of R2 (a) and RMSE (b). The test dataset includes 601 images for 6 classes. The vertical axis corresponds to the proportion of each class's pixels relative to the total pixels in the entire image.

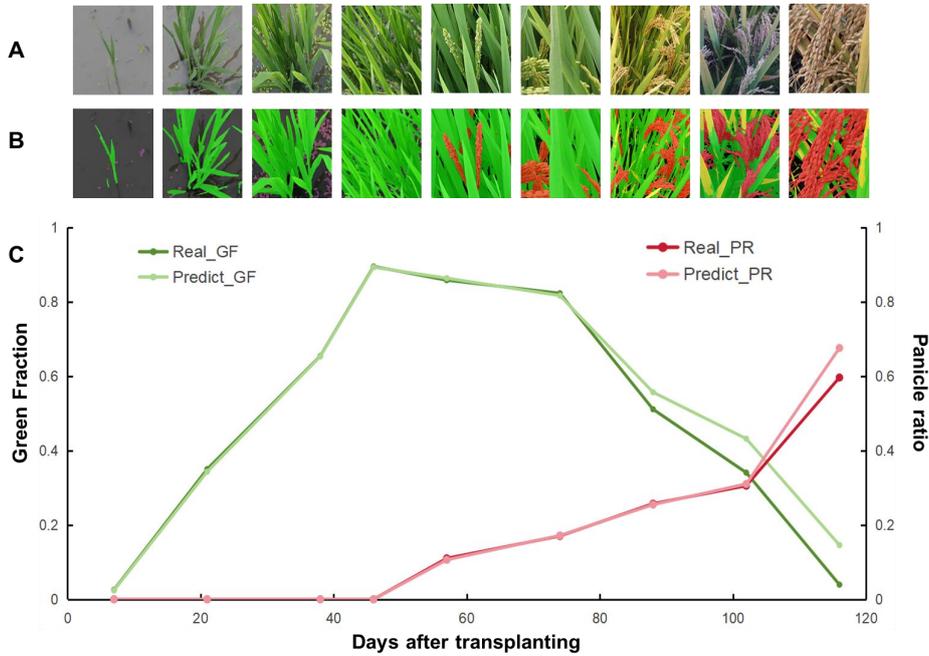

Figure 7. Dynamics of GF and Leaf to panicle ratio generated by time series images. (A) RGB images (B) Estimated (C) Dynamic of GF and panicle ratio.

## 4. Discussion

### 4.1. Potential contribution of the RiceSEG dataset

To the best of our knowledge, we established the first comprehensive multi-class rice semantic segmentation dataset, RiceSEG. We gathered nearly 50,000 high-resolution, ground-based images from five major rice-growing countries (China, Japan, India, the Philippines, and Tanzania), encompassing over 6,000 genotypes across all growth stages. From these original images, 3,078 representative samples were selected to form the RiceSEG dataset. Notably, the sub-dataset from China spans all major genotypes and rice-growing environments from the northeast to the south. The RiceSEG uniquely captures key rice crop organs, including the primary source organs—leaves (classified as green and senescent)—and the sink organ, the panicle. Unlike previous rice segmentation datasets, which were limited to binary segmentation of vegetation and background (Gao et al., 2023a), our dataset enables the development of advanced segmentation models to track the dynamics of these critical organs throughout the entire rice growth cycle (fig.6). By providing detailed time-series data on organ development, it offers insights that are unattainable through manual measurements, potentially unveiling new eco-physiological processes underlying crop adaptation to local environments and yield formation (Chang et al., 2023). Additionally, the dataset incorporates both aquatic and non-aquatic weeds, enabling simultaneous segmentation of weeds and rice crops. By facilitating accurate weed and crop differentiation, the dataset may play a crucial role in the development of advanced computer vision models for automated weed control, addressing the growing demand for precision agriculture solutions such as field robots (A. Wang et al., 2019). However, existing models encounter difficulties during the reproductive stage, when the canopy structures become more complex and multiple classes are involved. These findings highlight the importance of our dataset for developing specialized segmentation models for rice and other crops. Finally, through collaboration with international partners, we have expanded the dataset to include samples from 5 countries, representing diverse genotype-environment-management combinations. This broad representation ensures the robustness and scalability of the resulting segmentation models, enabling precise differentiation of fine phenotypic traits among hundreds or even thousands of genotypes for breeding programs.

### 4.2 Challenges in the rice image segmentation

Compared with other computer vision tasks, semantic segmentation in agriculture—particularly for rice—presents unique challenges. In the broader computer vision field, widely used datasets such as COCO and ADE20K typically encompass a larger number of categories and significantly more images than RiceSEG. However, these general-purpose datasets predominantly feature large objects with relatively planar surfaces, whereas crop images often contain dense, finely detailed structures—primarily leaves—characterized by numerous edges and complex spatial arrangements. This inherent complexity is further compounded by varying illumination within the canopy, where mutual shading and reduced light transmittance at greater canopy depths make it particularly difficult to segment leaves located near the bottom. In paddy rice fields, water surfaces introduce additional complications, including reflections and mirror-like effects that resemble vegetation, while submerged or partially submerged weeds add yet another layer of segmentation difficulty. Although a few existing datasets address crop image segmentation, their limited scope and categories do not fully capture the complexity of real-world agricultural settings. Consequently, our RiceSEG dataset offers distinct value for developing and validating specialized segmentation models tailored to rice and other plant species.

Because crop image segmentation datasets are both scarce and unique, current state-of-the-art methods are not fully optimized for the complexities inherent in rice imagery. Nonetheless, due to the robust feature-extraction capabilities of deep learning models, most tested architectures accurately classify dominant image components (e.g., background and green vegetation) at the pixel level. Beyond pixel-level performance, we also evaluated segmentation accuracy at the image scale, as many phenotypic trait estimations (e.g., the green vegetation fraction for Green Area Index, (J. Wang et al., 2020)) depend on organ-specific pixel fractions. Overall, image-level evaluations largely parallel pixel-level results but exhibit slight improvements, potentially due to compositional effects across each image. However, pronounced performance gaps remain in more challenging categories such as senescent vegetation, weeds, and duckweed. Transformer-based models (e.g., SegFormer and Mask2Former) demonstrate superiority in these domains, likely because their self-attention mechanisms capture long-distance dependencies and effectively handle intricate visual patterns (Dosovitskiy et al., 2020). By contrast, CNN-based architectures, which primarily extract local features, struggle to recognize the fine structures that require a more global contextual understanding (He et al., 2016). Moving forward, research could focus on further refining Transformer-based models to enhance segmentation performance in these nuanced categories.

### 4.3 Limitations of the dataset

We made significant efforts to collect rice images from the most representative rice-growing conditions. Nevertheless, our dataset still has limitations regarding its overall representativeness. For instance, in China, we gathered images from nearly all major rice-producing regions, capturing a wide range of genotype–environment–management combinations. In contrast, although we obtained an almost equivalent number of images in Japan, the Philippines, and India, their geographic and genotypic diversity is much narrower, potentially biasing the model towards Chinese conditions and reducing its generalizability elsewhere. Additionally in assembling each site's dataset, we included images spanning all growth stages to improve the model's ability to handle the entire crop cycle. Despite this, the annotated pixel counts across categories are imbalanced, particularly for senescent leaves, which constitute only 2.8% of annotated pixels. This imbalance may partly account for the relatively low segmentation accuracy observed for senescent leaves (fig.5). However, for images collected in natural environments, such pixel distribution is a normal representation of the natural world. Another factor could be the inherent ambiguity of annotating senescent leaves, especially those in lower canopy layers where shading is more pronounced. Furthermore, our current dataset does not include a detailed classification of weeds. To achieve more precise in-field weed management, a broader range of weed species is essential. Therefore, we are considering both collecting more field data and employing data generation techniques (Gao et al., 2023b, 2024b; Y. Li et al., 2023; Liu et al., 2019, 2021) to further diversify the dataset.

To facilitate distribution and track updates, we have provided detailed descriptions of the dataset at http://www.global-rice.com and http://www.phenix-lab.com. Unfortunately, open-access datasets remain scarce in plant phenotyping research. In contrast, the computer vision community has achieved rapid progress largely through shared resources that reduce redundant efforts and enhance efficiency. We encourage more researchers in plant phenotyping and digital agriculture to collectively foster an open-access culture. Such collaboration will expedite the development of robust deep learning algorithms for agricultural applications, ultimately making a greater impact on crop breeding and smart farming.

## Data Availability

The RiceSEG dataset is publicly available at http://www.global-rice.com/.

## Conflicts of Interest

The authors declare that there is no conflict of interest that is relevant to the content of this article.

## Authors' Contributions

Conceptualization, S.L., J.Z., and W.G.; Methodology, S.L., and W.G.; Software, J.Z., and H.W.; Validation, J.Z.; Formal

analysis, S.L., J.Z., and W.G.; Investigation, J.Z.. Y.G. and F.X.; Data curation, W.G., Y.K., T.N., M.B., P.R., K.K., H.L., Y.M., W.Y., W.L., J.W., F.Y., J.Z., X.H., Y.Y. and W.W.; Writing original draft preparation, J.Z.; Writing review and editing, J.Z., S.L., and W.G.; Supervision, S.L. and W.G.

## Funding

This work was supported by the National Key R&D Program of China (No. 2022YFD2300700, No. 2022YFE0116200), Young Scientists Fund of the National Natural Science Foundation of China (No. 42201437), the PhD Scientific Research and Innovation Foundation of The Education Department of Hainan Province Joint Project of Sanya Yazhou Bay Science and Technology City(No.HSPHDSRF-2024-09-001), the Japan Society for the Promotion of Science (No. 22KK0083), and the Sarabetsu Village "Endowed Chair for Field Phenomics" project in Hokkaido, Japan.

## Acknowledgment

We thank Professor Shirong Zhou from the State Key Laboratory of Crop Genetics & Germplasm Enhancement at Nanjing Agricultural University and Professor Lizhong Xiong from the National Key Laboratory of Crop Genetic Improvement at Huazhong Agricultural University for providing the data support.

# Reference

1. Bai, X., Liu, P., Cao, Z., Lu, H., Xiong, H., Yang, A., Cai, Z., Wang, J., & Yao, J. (2023). Rice Plant Counting, Locating, and Sizing Method Based on High-Throughput UAV RGB Images. *Plant Phenomics*, *5*, 0020. doi: 10.34133/plantphenomics.0020
2. Baret, F., Madec, S., Irfan, K., Lopez, J., Comar, A., Hemmerlé, M., Dutartre, D., Praud, S., & Tixier, M. H. (2018). Leaf-rolling in maize crops: from leaf scoring to canopy-level measurements for phenotyping. *Journal of Experimental Botany*, *69*(10), 2705–2716. doi: 10.1093/jxb/ery071
3. Cassman, K. G., & Harwood, R. R. (1995). The nature of agricultural systems: food security and environmental balance. *Food Policy*, *20*(5), 439–454. doi: 10.1016/0306-9192(95)00037-F
4. Chang, T.-G., Wei, Z.-W., Shi, Z., Xiao, Y., Zhao, H., Chang, S.-Q., Qu, M., Song, Q., Chen, F., Miao, F., & Zhu, X.-G. (2023). Bridging photosynthesis and crop yield formation with a mechanistic model of whole-plant carbon–nitrogen interaction. *In Silico Plants*, *5*(2), diad011. doi: 10.1093/insilicoplants/diad011
5. Chen, C., & Mcnairn, H. (2006). A neural network integrated approach for rice crop monitoring. *International Journal of Remote Sensing*, *27*(7), 1367–1393. doi: 10.1080/01431160500421507
6. Chen, L.-C., Zhu, Y., Papandreou, G., Schroff, F., & Adam, H. (2018). Encoder-Decoder with Atrous Separable Convolution for Semantic Image Segmentation. V. Ferrari, M. Hebert, C. Sminchisescu, & Y. Weiss (Eds.), Computer Vision – ECCV 2018 (pp. 833–851). Cham: Springer International Publishing. doi: 10.1007/978-3-030-01234-2_49
7. Cheng, B., Misra, I., Schwing, A. G., Kirillov, A., & Girdhar, R. (2022). *Masked-attention Mask Transformer for Universal Image Segmentation* (arXiv:2112.01527). arXiv. doi: 10.48550/arXiv.2112.01527
8. Cordts, M., Omran, M., Ramos, S., Rehfeld, T., Enzweiler, M., Benenson, R., Franke, U., Roth, S., & Schiele, B. (2016). The Cityscapes Dataset for Semantic Urban Scene Understanding. *2016 IEEE Conference on Computer Vision and Pattern Recognition (CVPR)*, 3213–3223. doi: 10.1109/CVPR.2016.350
9. David, E., Madec, S., Sadeghi-Tehran, P., Aasen, H., Zheng, B., Liu, S., Kirchgessner, N., Ishikawa, G., Nagasawa, K., Badhon, M. A., Pozniak, C., de Solan, B., Hund, A., Chapman, S. C., Baret, F., Stavness, I., & Guo, W. (2020). Global Wheat Head Detection (GWHD) Dataset: A Large and Diverse Dataset of High-Resolution RGB-Labelled Images to Develop and Benchmark Wheat Head Detection Methods. *Plant Phenomics*, *2020*. doi: 10.34133/2020/3521852
10. David, E., Serouart, M., Smith, D., Madec, S., Velumani, K., Liu, S., Wang, X., Pinto, F., Shafiee, S., Tahir, I. S. A., Tsujimoto, H., Nasuda, S., Zheng, B., Kirchgessner, N., Aasen, H., Hund, A., Sadhegi-Tehran, P., Nagasawa, K., Ishikawa, G., … Guo, W. (2021). Global Wheat Head Detection 2021: An Improved Dataset for Benchmarking Wheat Head Detection Methods. *Plant Phenomics*, *2021*. doi: 10.34133/2021/9846158
11. Dosovitskiy, A., Beyer, L., Kolesnikov, A., Weissenborn, D., Zhai, X., Unterthiner, T., Dehghani, M., Minderer, M., Heigold, G., Gelly, S., Uszkoreit, J., & Houlsby, N. (2020, October 2). *An Image is Worth 16x16 Words: Transformers for Image Recognition at Scale*. International Conference on Learning Representations. Retrieved from https://openreview.net/forum?id=YicbFdNTTy
12. Gao, Y., Li, L., Weiss, M., Guo, W., Shi, M., Lu, H., Jiang, R., Ding, Y., Nampally, T., Rajalakshmi, P., Baret, F., & Liu, S. (2024a). Bridging real and simulated data for cross-spatial- resolution vegetation segmentation with application to rice crops. *ISPRS Journal of Photogrammetry and Remote Sensing*, *218*, 133–150. doi: 10.1016/j.isprsjprs.2024.10.007
13. Gao, Y., Li, L., Weiss, M., Guo, W., Shi, M., Lu, H., Jiang, R., Ding, Y., Nampally, T., Rajalakshmi, P., Baret, F., & Liu, S. (2024b). Bridging real and simulated data for cross-spatial- resolution vegetation segmentation with application to rice crops. *ISPRS Journal of Photogrammetry and Remote Sensing*, *218*, 133–150. doi: 10.1016/j.isprsjprs.2024.10.007
14. Gao, Y., Li, Y., Jiang, R., Zhan, X., Lu, H., Guo, W., Yang, W., Ding, Y., & Liu, S. (2023a). Enhancing Green Fraction

Estimation in Rice and Wheat Crops: A Self-Supervised Deep Learning Semantic Segmentation Approach. *Plant Phenomics*, *5*, 0064. doi: 10.34133/plantphenomics.0064

15. Gao, Y., Li, Y., Jiang, R., Zhan, X., Lu, H., Guo, W., Yang, W., Ding, Y., & Liu, S. (2023b). Enhancing Green Fraction Estimation in Rice and Wheat Crops: A Self-Supervised Deep Learning Semantic Segmentation Approach. *Plant Phenomics*, *5*, 0064. doi: 10.34133/plantphenomics.0064
16. Garcia-Garcia, A., Orts-Escolano, S., Oprea, S., Villena-Martinez, V., & Garcia-Rodriguez, J. (2017, April 22). *A Review on Deep Learning Techniques Applied to Semantic Segmentation*. arXiv.Org. Retrieved from https://arxiv.org/abs/1704.06857v1
17. Godfray, H. C. J., Beddington, J. R., Crute, I. R., Haddad, L., Lawrence, D., Muir, J. F., Pretty, J., Robinson, S., Thomas, S. M., & Toulmin, C. (2010). Food Security: The Challenge of Feeding 9 Billion People. *Science*, *327*(5967), 812–818. doi: 10.1126/science.1185383
18. Haug, S., & Ostermann, J. (2015). A Crop/Weed Field Image Dataset for the Evaluation of Computer Vision Based Precision Agriculture Tasks. L. Agapito, M. M. Bronstein, & C. Rother (Eds.), Computer Vision - ECCV 2014 Workshops (pp. 105–116). Cham: Springer International Publishing. doi: 10.1007/978-3-319-16220-1_8
19. He, K., Zhang, X., Ren, S., & Sun, J. (2016). Deep Residual Learning for Image Recognition. *2016 IEEE Conference on Computer Vision and Pattern Recognition (CVPR)*, 770–778. doi: 10.1109/CVPR.2016.90
20. Jia, Z., Chen, J., Xu, X., Kheir, J., Hu, J., Xiao, H., Peng, S., Hu, X. S., Chen, D., & Shi, Y. (2023). The importance of resource awareness in artificial intelligence for healthcare. *Nature Machine Intelligence*, *5*(7), 687–698. doi: 10.1038/s42256-023-00670-0
21. Jin, Z., Shah, T., Zhang, L., Liu, H., Peng, S., & Nie, L. (2020). Effect of straw returning on soil organic carbon in rice–wheat rotation system: A review. *Food and Energy Security*, *9*(2), e200. doi: 10.1002/fes3.200
22. Kirillov, A., Mintun, E., Ravi, N., Mao, H., Rolland, C., Gustafson, L., Xiao, T., Whitehead, S., Berg, A. C., Lo, W.-Y., Dollár, P., & Girshick, R. (2023). *Segment Anything* (arXiv:2304.02643). arXiv. Retrieved from http://arxiv.org/abs/2304.02643
23. Li, Y., Zhan, X., Liu, S., Lu, H., Jiang, R., Guo, W., Chapman, S., Ge, Y., Solan, B., Ding, Y., & Baret, F. (2023). Self-Supervised Plant Phenotyping by Combining Domain Adaptation with 3D Plant Model Simulations: Application to Wheat Leaf Counting at Seedling Stage. *Plant Phenomics*, *5*, 0041. doi: 10.34133/plantphenomics.0041
24. Li, Z., Guo, R., Li, M., Chen, Y., & Li, G. (2020). A review of computer vision technologies for plant phenotyping. *Computers and Electronics in Agriculture*, *176*, 105672. doi: 10.1016/j.compag.2020.105672
25. Liu, S., Baret, F., Abichou, M., Manceau, L., Andrieu, B., Weiss, M., & Martre, P. (2021). Importance of the description of light interception in crop growth models. *Plant Physiology*, *186*(2), 977–997. doi: 10.1093/plphys/kiab113
26. Liu, S., Martre, P., Buis, S., Abichou, M., Andrieu, B., & Baret, F. (2019). Estimation of Plant and Canopy Architectural Traits Using the Digital Plant Phenotyping Platform1 [OPEN]. *Plant Physiology*, *181*(3), 881–890. doi: 10.1104/pp.19.00554
27. Long, J., Shelhamer, E., & Darrell, T. (2015). *Fully Convolutional Networks for Semantic Segmentation* (arXiv:1411.4038). arXiv. doi: 10.48550/arXiv.1411.4038
28. Madec, S., Irfan, K., Velumani, K., Baret, F., David, E., Daubige, G., Samatan, L. B., Serouart, M., Smith, D., James, C., Camacho, F., Guo, W., De Solan, B., Chapman, S. C., & Weiss, M. (2023). VegAnn, Vegetation Annotation of multi-crop RGB images acquired under diverse conditions for segmentation. *Scientific Data*, *10*(1), 302. doi: 10.1038/s41597-023-02098-y
29. Madec, S., Jin, X., Lu, H., De Solan, B., Liu, S., Duyme, F., Heritier, E., & Baret, F. (2019). Ear density estimation from high resolution RGB imagery using deep learning technique. *Agricultural and Forest Meteorology*, *264*, 225–234. doi: 10.1016/j.agrformet.2018.10.013

30. Mandal, D., Kumar, V., Bhattacharya, A., Rao, Y. S., Siqueira, P., & Bera, S. (2018). Sen4Rice: A Processing Chain for Differentiating Early and Late Transplanted Rice Using Time-Series Sentinel-1 SAR Data With Google Earth Engine. *IEEE Geoscience and Remote Sensing Letters*, *15*(12), 1947–1951. doi: 10.1109/LGRS.2018.2865816
31. Maohua, W. (2001). Possible adoption of precision agriculture for developing countries at the threshold of the new millennium. *Computers and Electronics in Agriculture*, *30*(1), 45–50. doi: 10.1016/S0168-1699(00)00154-X
32. Mermut, A. R., & Eswaran, H. (2001). Some major developments in soil science since the mid-1960s. *Geoderma*, *100*(3), 403–426. doi: 10.1016/S0016-7061(01)00030-1
33. Mortensen, A. K., Skovsen, S., Karstoft, H., & Gislum, R. (2019). The Oil Radish Growth Dataset for Semantic Segmentation and Yield Estimation. *2019 IEEE/CVF Conference on Computer Vision and Pattern Recognition Workshops (CVPRW)*, 2703–2710. doi: 10.1109/CVPRW.2019.00328
34. Prajapati, H. B., Shah, J. P., & Dabhi, V. K. (2017). Detection and classification of rice plant diseases. *Intelligent Decision Technologies*, *11*(3), 357–373. doi: 10.3233/IDT-170301
35. Russell, B. C., Torralba, A., Murphy, K. P., & Freeman, W. T. (2008). LabelMe: A Database and Web-Based Tool for Image Annotation. *International Journal of Computer Vision*, *77*(1), 157–173. doi: 10.1007/s11263-007-0090-8
36. Scharr, H., Minervini, M., Fischbach, A., & Tsaftaris, S. (2014). *Annotated Image Datasets of Rosette Plants*.
37. Serouart, M., Madec, S., David, E., Velumani, K., Lopez Lozano, R., Weiss, M., & Baret, F. (2022). SegVeg: Segmenting RGB Images into Green and Senescent Vegetation by Combining Deep and Shallow Methods. *Plant Phenomics*, *2022*. doi: 10.34133/2022/9803570
38. Shelhamer, E., Long, J., & Darrell, T. (2017). Fully Convolutional Networks for Semantic Segmentation. *IEEE Transactions on Pattern Analysis and Machine Intelligence*, *39*(4), 640–651. doi: 10.1109/TPAMI.2016.2572683
39. Shouyang, L., Shichao, J., Qinghua, G., Yan, Z., & Fred, B. (2020). An algorithm for estimating field wheat canopy light interception based on Digital Plant Phenotyping Platform. *Smart Agriculture*, *2*(1), 87. doi: 10.12133/j.smartag.2020.2.1.202002-SA004
40. Tangseng, P., Wu, Z., & Yamaguchi, K. (2017). *Looking at Outfit to Parse Clothing* (arXiv:1703.01386). arXiv. doi: 10.48550/arXiv.1703.01386
41. Wang, A., Zhang, W., & Wei, X. (2019). A review on weed detection using ground-based machine vision and image processing techniques. *Computers and Electronics in Agriculture*, *158*, 226–240. doi: 10.1016/j.compag.2019.02.005
42. Wang, H., Lyu, S., & Ren, Y. (2021). Paddy Rice Imagery Dataset for Panicle Segmentation. *Agronomy*, *11*(8), 1542. doi: 10.3390/agronomy11081542
43. Wang, J., Lopez-Lozano, R., Weiss, M., Buis, S., Li, W., Zhang, J., & Baret, F. (2020). *Estimating Green Area Index (GAI) from radiative transfer model: application of the Bayesian theory to account for crop-specificities*. *2020*, B009-10.
44. Weyler, J., Magistri, F., Marks, E., Chong, Y. L., Sodano, M., Roggiolani, G., Chebrolu, N., Stachniss, C., & Behley, J. (2023). *PhenoBench -- A Large Dataset and Benchmarks for Semantic Image Interpretation in the Agricultural Domain* (arXiv:2306.04557). arXiv. doi: 10.48550/arXiv.2306.04557
45. Wu, X., Zhan, C., Lai, Y.-K., Cheng, M.-M., & Yang, J. (2019). IP102: A Large-Scale Benchmark Dataset for Insect Pest Recognition. *2019 IEEE/CVF Conference on Computer Vision and Pattern Recognition (CVPR)*, 8779–8788. doi: 10.1109/CVPR.2019.00899
46. Xie, E., Wang, W., Yu, Z., Anandkumar, A., Alvarez, J. M., & Luo, P. (2021). SegFormer: Simple and Efficient Design for Semantic Segmentation with Transformers. *Advances in Neural Information Processing Systems*, *34*, 12077–12090. Retrieved from https://proceedings.neurips.cc/paper/2021/hash/64f1f27bf1b4ec22924fd0acb550c235-Abstract.html
47. Yandun Narvaez, F., Reina, G., Torres-Torriti, M., Kantor, G., & Cheein, F. A. (2017). A Survey of Ranging and Imaging

Techniques for Precision Agriculture Phenotyping. *IEEE/ASME Transactions on Mechatronics*, *22*(6), 2428–2439. doi: 10.1109/TMECH.2017.2760866

48. Zhang, W., Pang, J., Chen, K., & Loy, C. C. (2024). K-Net: towards unified image segmentation. *Proceedings of the 35th International Conference on Neural Information Processing Systems*, 10326–10338.

49. Zhao, H., Shi, J., Qi, X., Wang, X., & Jia, J. (2017). *Pyramid Scene Parsing Network*. 6230–6239. doi: 10.1109/CVPR.2017.660

50. Zhao, Z., Wang, C., Yu, X., Tian, Y., Wang, W., Zhang, Y., Bai, W., Yang, N., Zhang, T., Zheng, H., Wang, Q., Lu, J., Lei, D., He, X., Chen, K., Gao, J., Liu, X., Liu, S., Jiang, L., … Wan, J. (2022). Auxin regulates source-sink carbohydrate partitioning and reproductive organ development in rice. *Proceedings of the National Academy of Sciences*, *119*(36), e2121671119. doi: 10.1073/pnas.2121671119